\documentclass[runningheads]{llncs}

\usepackage[T1]{fontenc}
\usepackage{times}

\usepackage{graphicx}
\usepackage[breaklinks,colorlinks,allcolors=blue]{hyperref}
\usepackage{color}

\usepackage{amsmath}
\usepackage{amssymb}  %
\usepackage{amsfonts}
\usepackage[font=small,skip=2pt]{caption}
\usepackage{booktabs} %
\usepackage[normalem]{ulem} %
\usepackage{array} %
\usepackage{lipsum}

\title{Scheduling the Off-Diagonal Weingarten Loss of Neural SDFs for CAD Models}

\author{
    Haotian Yin \and Przemyslaw Musialski
}
\authorrunning{H. Yin and P. Musialski}

\institute{
    New Jersey Institute of Technology, United States \\
    \email{\{hy9, przem\}@njit.edu}
}

\begin{document}

\maketitle

\begin{abstract}
Neural signed distance functions (SDFs) have become a powerful representation for geometric reconstruction from point clouds, yet they often require both gradient- and curvature-based regularization to suppress spurious warp and preserve structural fidelity. FlatCAD introduced the Off-Diagonal Weingarten (ODW) loss as an efficient second-order prior for CAD surfaces, approximating full-Hessian regularization at roughly half the computational cost. However, FlatCAD applies a fixed ODW weight throughout training, which is suboptimal: strong regularization stabilizes early optimization but suppresses detail recovery in later stages. We present scheduling strategies for the ODW loss that assign a high initial weight to stabilize optimization and progressively decay it to permit fine-scale refinement. We investigate constant, linear, quintic, and step interpolation schedules, as well as an increasing warm-up variant. Experiments on the ABC CAD dataset demonstrate that time-varying schedules consistently outperform fixed weights. Our method achieves up to a 35\% improvement in Chamfer Distance over the FlatCAD baseline, establishing scheduling as a simple yet effective extension of curvature regularization for robust CAD reconstruction. 
\end{abstract}

\section{Introduction}

Neural implicit fields, in particular signed distance functions (SDFs), have become a central representation in 3D vision and graphics~\cite{Xie2021NeuralFI}. They encode geometry as the zero-level set of a neural network, which makes them continuous, differentiable, and resolution-independent.  
As a result, they have been widely adopted for geometric reconstruction from point clouds and scans, and as a backbone for applications such as meshing, rendering, and shape analysis.~\cite{Park2019DeepSDFLC,Mescheder2018OccupancyNL,Sitzmann2020ImplicitNR,Mller2022InstantNG}

For computer-aided design (CAD) data, implicit fields face a specific challenge. CAD parts are largely composed of simple developable primitives—planes, cylinders, cones—and their blends, joined by sharp feature curves. Training with only point-wise losses may produce shapes that interpolate the samples but violate these structural constraints, leading to warped or curved regions where the target is flat or cylindrical. To avoid such artifacts, additional regularization terms are required to enforce geometric faithfulness during learning.

Several approaches have introduced curvature-based priors. The Implicit Geometric Regularizer (IGR) enforces the Eikonal property~\cite{Gropp2020ImplicitGR}, DiGS penalizes divergence~\cite{BenShabat2021DiGSD}, Neural-Singular-Hessian (NSH) constrains the Hessian rank~\cite{Wang2023NeuralSingularHessianIN}, and NeurCADRecon (NCR) penalizes Gaussian curvature across a surface shell~\cite{Dong2024NeurCADReconNR}. While effective, such methods require explicit or repeated access to the full Hessian of the network, incurring prohibitive cost in memory and runtime.

FlatCAD~\cite{Yin2025FlatCAD} recently proposed a more efficient alternative: the {Off-Diagonal Weingarten} (ODW) loss. Instead of evaluating all second-order terms, ODW regularizes only the mixed entry of the shape operator, which encodes the gap between principal curvatures. This loss can be estimated with a single Hessian--vector product or a finite-difference stencil, matching the accuracy of full-Hessian methods while reducing training cost by a factor of two. FlatCAD demonstrated that ODW performs well on reconstructions on CAD benchmarks with faster convergence and less GPU memory usage. Yin et. al generalized the finite-differences approach to other  operators~\cite{Yin2025_finitedif}. 

A major limitation of that work is that the curvature weight $\lambda_{\text{ODW}}$ is fixed throughout training. FlatCAD showed that performance depends sensitively on this choice—if too weak, residual warp persists, while if too strong, cylinders and cones collapse toward planar or spherical limits. A constant weight is unlikely to be optimal across the full optimization process, where early iterations benefit from strong regularization while later stages demand flexibility to match fine detail.

This paper addresses that gap. We present the first systematic study of \emph{ODW scheduling}, investigating how time-dependent weighting strategies for $\lambda_{\text{ODW}}$ affect reconstruction accuracy, convergence, and robustness. We evaluate constant baselines, as well as several annealing schemes including linear decay, quintic (smooth polynomial) decay, and stepwise schedules. In addition, we contrast decreasing (strong–start/decay) and increasing (warm-up) variants to isolate the effect of early versus late regularization. Experiments on the ABC CAD dataset~\cite{Koch_2019_CVPR} demonstrate that appropriate scheduling can accelerate convergence and consistently improve reconstruction fidelity without incurring additional runtime cost.

The contribution of this work is practical but clear: it establishes annealing of the ODW weight as an effective design dimension for curvature regularization in neural implicit SDFs, extending the utility of ODW loss toward scalable and reliable CAD reconstruction. Our experiments show consistent improvements over the constant-weight baseline in both quantitative metrics and qualitative surface quality.

\section{Related Work}

\emph{Neural Implicit Representations.}
Implicit neural fields have emerged as a flexible alternative to explicit surface models~\cite{Xie2021NeuralFI}. Early works such as DeepSDF~\cite{Park2019DeepSDFLC} and Occupancy Networks~\cite{Mescheder2018OccupancyNL} showed that multilayer perceptrons can represent watertight surfaces with continuous evaluation and differentiation. Subsequent advances like SIREN~\cite{Sitzmann2020ImplicitNR} and Instant-NGP~\cite{Mller2022InstantNG} improved fidelity and training speed, making implicit models a standard tool for geometry learning.

\noindent
\emph{Curvature Regularization for Implicit Surfaces.}
Despite their expressiveness, vanilla implicit fields often produce surfaces with irregular curvature. Several geometric priors have been proposed to address this. The Implicit Geometric Regularizer (IGR)~\cite{Gropp2020ImplicitGR} enforces the Eikonal constraint, while DiGS~\cite{BenShabat2021DiGSD} adds a divergence penalty to stabilize training. Neural-Singular-Hessian (NSH)~\cite{Wang2023NeuralSingularHessianIN} encourages rank-deficient Hessians, suppressing spurious curvature oscillations. NeurCADRecon (NCR)~\cite{Dong2024NeurCADReconNR} targets CAD data explicitly by minimizing Gaussian curvature, thereby promoting developable patches. FlatCAD~\cite{Yin2025FlatCAD} introduced the Off-Diagonal Weingarten (ODW) loss, which regularizes only the curvature gap between principal curvatures. This proxy matches the accuracy of full-Hessian penalties while reducing runtime and memory.

\noindent
\emph{Limitations of Fixed Regularization.}
All of the above methods employ fixed weighting coefficients for their regularizers. Empirical studies consistently show that performance depends strongly on this choice: too little regularization leaves surfaces noisy, while too much suppresses genuine curvature such as cylinders or cones. FlatCAD ablated a wide range of $\lambda_{\text{ODW}}$ values and confirmed that reconstruction quality varies with the weight, but did not explore changing it during training. Thus, prior work establishes the need for curvature priors but treats their balance with data fidelity as static.

\noindent
\emph{Curriculum Learning and Scheduling.}
In contrast, research on curriculum and multi-stage training demonstrates the value of dynamic weighting. Curriculum DeepSDF~\cite{cirsdf} gradually increased the weight of difficult samples, yielding reconstructions superior to any fixed setting. Neuralangelo~\cite{neuralangelo} adopted a coarse-to-fine strategy: strong smoothing early, progressively reduced to recover fine detail. These examples show that time-varying objectives can produce better optimization trajectories and higher-fidelity results than static trade-offs.

\noindent
\emph{Dynamic Loss Balancing.}
A parallel line of work in multi-task learning develops adaptive methods for balancing competing objectives. Kendall et al.~\cite{Multi-task_learning} proposed uncertainty-based weighting that learns each loss coefficient automatically. GradNorm~\cite{gradnorm} equalizes training rates by adjusting weights so that all objectives contribute comparable gradient magnitudes. Guo et al.~\cite{guo2018dynamic} introduced dynamic task prioritization, emphasizing harder tasks as training evolves. Although developed outside geometry, these approaches directly apply: the data term and curvature term can be treated as two tasks whose relative influence should adapt to training dynamics.

\noindent
\emph{Summary.}
Prior work demonstrates that (i) curvature regularization is essential for faithful implicit surfaces, but (ii) the effect of regularizers is highly sensitive to their weighting, and (iii) scheduling or adaptive weighting can outperform fixed settings in related contexts. To our knowledge, no prior study has systematically investigated scheduling strategies for curvature regularization in neural SDFs. Our work addresses this gap by evaluating a range of schedules for the ODW loss and analyzing their effect on accuracy, stability, and efficiency in CAD reconstruction.

\section{Method}
This section reviews the geometric background and introduces our time-varying ODW weighting scheme. %

\subsection{Background: SDFs and Curvature}
\begin{figure*}[t]
    \centering
    \includegraphics[width=0.99\linewidth]{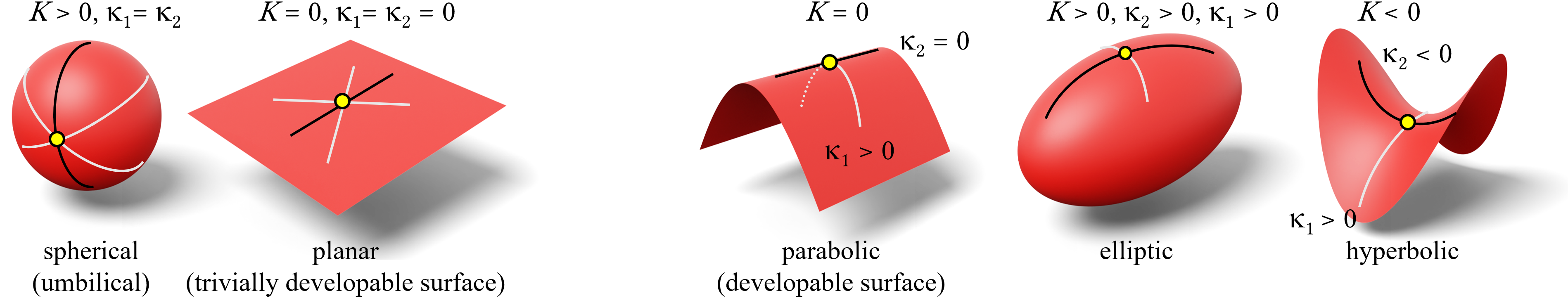}
    \caption{Principal curvatures $\kappa_1$ and $\kappa_2$ are the eigenvalues of the Weingarten map $S$. Their product gives the Gaussian curvature $K=\kappa_1 \kappa_2 = \det S$. Because $S$ is self-adjoint with respect to the first fundamental form, its eigenvectors (the principal directions) are orthogonal in the surface metric whenever the eigenvalues are distinct. At an umbilic point (e.g. on a sphere) the entire two-dimensional tangent plane is the eigenspace, so no unique pair of principal directions exists. For a plane the second fundamental form vanishes, i.e., $S=0$.}
    \label{fig:curvature}
\end{figure*}

A signed distance field (SDF) is a scalar function $f:\mathbb{R}^3 \to \mathbb{R}$ whose zero level set $\{x \mid f(x)=0\}$ defines a surface. 
By construction, a true SDF satisfies $\|\nabla f\|=1$ in a neighborhood of the surface, so the normalized gradient $n=\nabla f/\|\nabla f\|$ provides a consistent surface normal. 
The second derivatives of $f$ capture curvature: the Hessian $H_f$ encodes how the gradient field bends, and projecting $H_f$ into a local tangent basis $(u,v)$ yields the shape operator
\[
S = 
\begin{bmatrix}
u^\top H_f u & u^\top H_f v \\
v^\top H_f u & v^\top H_f v
\end{bmatrix}. 
\]
That operator (Weingarten map) takes any  tangent vector and returns how fast the surface normal rotates in that direction, i.e., the local bending.  
Its eigenvalues are the \emph{principal curvatures} $\kappa_1,\kappa_2$; their product gives the Gaussian curvature $K=\kappa_1\kappa_2=\det S$. Together, these values distinguish spherical, planar, parabolic, elliptic, and hyperbolic regimes as illustrated in Fig.~\ref{fig:curvature}. 
Consequently, controlling curvature during learning has an important impact on the neighborhood of a sample~\cite{Yin2025FlatCAD}: without regularization, neural SDFs may satisfy pointwise data terms but introduce spurious warp and oscillations.

\subsection{Surface Reconstruction and Off-Diagonal Weingarten Loss} 
To keep flat regions planar, which is common in CAD-type geometric data, 
Yin et al.~\cite{Yin2025FlatCAD} proposed the Off-Diagonal Weingarten (ODW) loss as a lightweight curvature proxy. 
For any rotated tangent frame, they show that the off-diagonal entry of $S$ is
\[
S_{12}(\theta) = \tfrac{1}{2}(\kappa_2-\kappa_1)\sin 2\theta, 
\]
which vanishes if and only if $\kappa_1=\kappa_2$. 
Thus, penalizing $|S_{12}|$ suppresses the curvature gap while allowing each curvature to follow the data. 
This regularizer flattens parabolic and hyperbolic patches and rounds elliptic ones uniformly, improving geometric faithfulness without computing full Gaussian curvature (cf. Figure~\ref{fig:curvature}).  
FlatCAD showed that minimizing ODW matches in results the accuracy of full-Hessian baselines while roughly halving runtime and memory.

\subsection{Baseline: FlatCAD Losses}

Let $f:\mathbb{R}^3 \to \mathbb{R}$ denote the signed distance field predicted by the network. 
We follow FlatCAD~\cite{Yin2025FlatCAD} and combine standard reconstruction terms with curvature regularization. 
Given on-surface samples $\mathcal{X}_{\mathrm{man}}=\{x_i\}_{i=1}^N$ (with part labels available on a subset $\mathcal{X}_{\mathrm{lab}} \subseteq \mathcal{X}_{\mathrm{man}}$), 
off-surface samples $\mathcal{X}_{\mathrm{non}}=\{y_j\}_{j=1}^M$, 
and near-surface shell samples $\Omega=\{p_\ell\}_{\ell=1}^L$, 
the total loss is
\begin{equation}
\mathcal{L}_{\mathrm{total}}
= \lambda_{\mathrm{DM}} \mathcal{L}_{\mathrm{DM}}
+ \lambda_{\mathrm{DNM}} \mathcal{L}_{\mathrm{DNM}}
+ \lambda_{\mathrm{EIK}} \mathcal{L}_{\mathrm{EIK}}
+ \lambda_{\mathrm{ODW}} \mathcal{L}_{\mathrm{ODW}} .
\end{equation}
where $\lambda$ denotes the weighting coefficients for the corresponding loss terms.

\paragraph{Manifold (Dirichlet) loss.}
We constrain on-surface samples, sampled directly from the input point cloud, to lie on the zero level set of the SDF~\cite{Atzmon2019SALSA}:
\begin{equation}
\mathcal{L}_{\mathrm{DM}}
= \frac{1}{N} \sum_{x \in \mathcal{X}_{\mathrm{man}}} \bigl| f(x) \bigr|.
\end{equation}
where $\mathcal{X}{\mathrm{man}}$ denotes the set of $N$ surface samples.
This term enforces geometric fidelity between the reconstructed implicit surface and the observed data.

\paragraph{Non-manifold (sign-agnostic) loss.}
To avoid spurious zero crossings far from the object, we sample $M$ points $\mathcal{X}{\mathrm{non}}$ uniformly within the normalized bounding box of the shape and apply a rapidly decaying~\cite{Atzmon2020SALSA}:
\begin{equation}
\mathcal{L}_{\mathrm{DNM}}
= \frac{1}{M} \sum_{y \in \mathcal{X}_{\mathrm{non}}} 
   \exp\!\bigl(-\alpha\,|f(y)|\bigr).
\end{equation}
where $\alpha = 100$ controls the decay rate.
This regularization encourages off-surface regions to maintain a consistent sign and prevents undesired zero-level artifacts.

\paragraph{Eikonal loss.}
Following IGR~\cite{Gropp2020ImplicitGR}, we enforce the SDF property by penalizing deviations of the gradient norm from unity $\|\nabla f\|_2 = 1$:
\begin{equation}
\mathcal{L}_{\mathrm{EIK}}
= \frac{1}{K_e} \sum_{z \in \mathcal{Z}}
  \Bigl( \|\nabla f(z)\|_2^2 - 1 \Bigr)^2 ,
\end{equation}
where $\mathcal{Z}$ is the union of the on-surface samples $\mathcal{X}{\mathrm{man}}$ and additional near-surface samples randomly generated in the surrounding space.

\paragraph{Off-diagonal Weingarten loss.}
FlatCAD introduces the Off-Diagonal Weingarten (ODW) loss to suppress curvature differences. 
For each $p \in \Omega$, let $n = \nabla f(p)/\|\nabla f(p)\|_2$ be the unit normal and $(u,v)$ any orthonormal tangent frame. 
The off-diagonal entry of the Weingarten map is
\begin{equation*}
S_{12}(p) = \frac{u^\top H_f(p)\,v}{\|\nabla f(p)\|_2},
\end{equation*}
where $H_f$ is the Hessian of $f$. 
The ODW loss penalizes its magnitude:
\begin{equation}
\mathcal{L}_{\mathrm{ODW}}
= \frac{1}{L} \sum_{p \in \Omega} \bigl| S_{12}(p) \bigr| .
\end{equation}
In practice, $S_{12}(p)$ is evaluated on the shell $\Omega$ for numerical stability using either a Hessian–vector product or a finite-difference stencil~\cite{Yin2025FlatCAD}. 
In that baseline setting, the weight $\lambda_{\mathrm{ODW}}$ remains constant and is hence {the subject} of our following investigations.

\subsection{Scheduling the ODW Weight}
In many implicit surface learning frameworks, such as {FlatCAD}, the weights assigned to different loss terms remain fixed throughout training. This implicitly assumes that all constraints—data fidelity, first-order Eikonal regularization~\cite{Gropp2020ImplicitGR}, and higher-order curvature terms—are equally critical at every stage of optimization. However, recent theoretical analyses challenge this assumption. In particular, Yang et al.~\cite{StEik} demonstrate that the widely used Eikonal loss, when viewed in the continuum limit, can induce an unstable partial differential equation (PDE) flow, causing the optimization to oscillate or converge to spurious local minima that obscure geometric detail. Conversely, higher-order constraints such as off-diagonal Weingarten (ODW) regularization provide a stabilizing effect by suppressing curvature irregularities, yet excessive weight on these terms may over-regularize the surface, driving the reconstruction toward overly smooth solutions and diminishing fidelity to the input. Similar annealing principles have been observed in recent methods such as Neural-Singular-Hessian, which emphasize second-order regularization strongly at the beginning to suppress ghost geometry, and gradually relax it to recover fine-scale details.

Motivated by these insights, we propose an annealing framework for the ODW regularization weight. Our strategy adopts a strong-start schedule: the ODW weight is set high during the early iterations to provide a global stabilizing prior that biases the optimization toward flat or developable geometries—an assumption consistent with the structural characteristics of CAD models. This suppresses unstable curvature artifacts and steers the optimization away from poor local minima. As training progresses, we gradually reduce the ODW weight, thereby relaxing the regularization and allowing the network to refine geometric details without being dominated by second-order smoothing. This staged strategy balances stability and fidelity: first constraining the solution to remain well-posed, then progressively freeing it to capture fine-scale structure.

\subsection{Scheduling Strategies}
To implement the proposed strong–start framework, we parameterize the ODW weight $\lambda_{\mathrm{ODW}}(t)$ as a piecewise interpolation between a set of user-specified control points, where $t \in [0,1]$ denotes the normalized training progress. We experiment with four families of schedules:

\paragraph{\textbf{Constant (baseline)}}  
As adopted in {FlatCAD}, the weight remains fixed throughout training, i.e.,
\[
\lambda_{\mathrm{ODW}}(t) = \lambda_0,
\]
which implicitly assumes equal importance of the ODW loss at all training stages.

\paragraph{\textbf{Linear scheduling}}  
A piecewise linear interpolation decreases or increases the weight at a constant rate, the most common annealing strategy in practice. For a segment spanning $[s,e]$, the weight evolves as
\[
\lambda_{\mathrm{ODW}}(t) = w_0 + (w_e - w_0)\,\frac{t-s}{e-s}, 
\qquad s \leq t \leq e,
\]
where $w_0$ and $w_e$ denote the start and end weights, respectively.

\paragraph{\textbf{Quintic scheduling}}  
To avoid abrupt changes, we employ a smooth fifth-order polynomial that remains nearly constant at first, decays rapidly around the middle of the interval, and then gradually flattens toward the end:
\[
\lambda_{\mathrm{ODW}}(t) = w_0 + (w_e - w_0)\Bigl(1 - \Bigl(1 - \tfrac{t-s}{e-s}\Bigr)^5\Bigr), 
\qquad s \leq t \leq e.
\]
This quintic easing provides a gentler transition and stabilizes training by preventing sudden shocks in the loss landscape.

\paragraph{\textbf{Step scheduling}}  
As a limiting case, the weight undergoes a discontinuous jump at a designated threshold $s$:
\[
\lambda_{\mathrm{ODW}}(t) = 
\begin{cases} 
w_0, & 0 \leq t < s_e, \\ 
w_e, & s_e \leq t < s_{e+1}, 
\end{cases}
\]
creating an instantaneous shift in the constraint strength. While simple to implement, such “shock” schedules may induce optimization transients.

\begin{figure*}[t]
    \centering
    \includegraphics[width=0.99\linewidth]{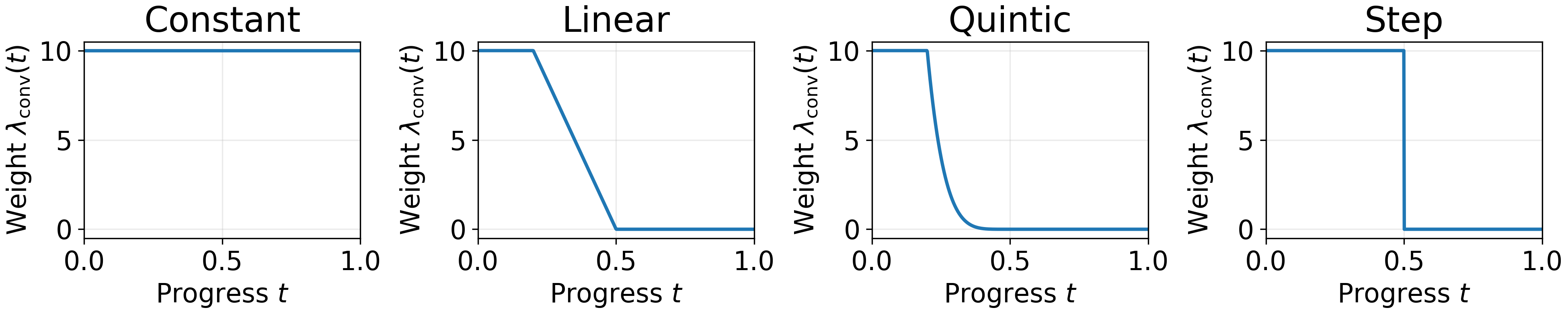}
    \caption{Scheduling strategies for the ODW weight $\lambda_{\mathrm{ODW}}(t)$.
From left to right: \textbf{Constant} (FlatCAD baseline), \textbf{Linear} (piecewise linear ramp),
\textbf{Quintic} (fifth-order polynomial interpolation), and \textbf{Step} (discontinuous jump).
All schedules share the same control points \big($[(0,10),\,(0.2,10),\,(0.5,0.001),\,(1.0,0)]$, and normalized time $t\!\in\![0,1]$, $n_{\text{iter}}{=}10{,}000$\big).
The apparent kink at $t=0.2$ in the quintic curve arises because the weight is held constant before 0.2 and only begins quintic interpolation afterward.  
}
    \label{fig:schedules}
\end{figure*}

\medskip
Together, these schedules offer a spectrum of trade-offs between stability and adaptivity. In practice, the linear and quintic interpolations produce the most stable optimization trajectories, while the step schedule serves as an ablation baseline.

\subsection{Scheduling implementation}
In practice, we implement the annealing schedule by specifying a small set of keypoints $(s_i, w_i)$, where each $s_i \in [0,1]$ denotes a normalized training progress (iteration ratio) and $w_i$ is the corresponding target weight. Between two consecutive keypoints $(s_i,w_i)$ and $(s_{i+1},w_{i+1})$, the interpolation rule is chosen according to the selected scheduling policy—linear, quintic (fifth-order polynomial), or step—yielding a piecewise-defined function $\lambda_{\mathrm{ODW}}(t)$ that is continuous in value and, in the case of linear or quintic interpolation, smoothly varying across training. During optimization, the current training iteration $t$ is normalized by the total number of iterations, the enclosing segment $[s_i,s_{i+1}]$ is located, and the interpolated weight is computed on-the-fly. This modular design makes the schedule easy to configure via a simple parameter list while ensuring reproducibility across experiments.

In our experiments, we instantiate the schedule with keypoints $(s_0,w_0)=(0,10)$, $(s_1,w_1)=(0.2,10)$, $(s_2,w_2)=(0.5,0.001)$, $(s_3,w_3)=(1.0,0)$. Concretely, the weight remains fixed at $10$ for the first $20\%$ of training, decays to $0.001$ by the halfway point, and finally converges to $0$ at the end of training. Thus, for $t \in [0,0.2]$, we have $\lambda_{\mathrm{ODW}}(t)=10$; for $t\in[0.2,0.5]$, the value is interpolated between $10$ and $0.001$; and for $t \in [0.5,1.0]$, it continues to decay toward zero. Between any two consecutive anchors $(s_i,w_i)$ and $(s_{i+1},w_{i+1})$, the interpolation rule is governed by the chosen scheduling policy—linear, quintic (fifth-order polynomial), or step—yielding a piecewise-defined function $\lambda_{\mathrm{ODW}}(t)$ that is continuous in value and, in the case of linear or quintic schedules, smoothly varying across training (cf. Figure~\ref{fig:schedules}).

\section{Experiments and Results}\label{sec:results}

In this section, we evaluate our proposed ODW weight–scheduling strategies. %

\subsection{Experimental Setup}

Our experiments are conducted on subsets of the ABC dataset~\cite{Koch_2019_CVPR}, consisting of 25 models. The dataset contains a pseudo-random collection of models with individual file sizes of approximately 1 MB, selected to ensure clean topology and clearly defined geometric features. For every mesh, we generate an input point cloud by uniformly sampling 30,000 surface points, thereby standardizing the input distribution across all methods. During training, 20,000 points are randomly drawn from this pool at each iteration, and an additional 20,000 off-surface samples are produced through uniform spatial sampling within the mesh bounding volume.

\subsection{Methods and Experimental Setup}
We evaluate our proposed scheduling strategies (Constant, Linear, Quintic, and Step interpolation) under a unified experimental framework to ensure fairness. All variants share the same backbone architecture: a SIREN-based MLP~\cite{Sitzmann2020ImplicitNR} with four hidden layers of 256 units and sine activations, initialized using the standard SIREN scheme. Training is performed with the Adam optimizer~\cite{kingma2017adam} at a fixed learning rate of $5\times10^{-5}$ for up to 10,000 iterations. An early-stopping criterion is triggered when the Chamfer Distance fails to improve for 1,500 consecutive iterations. Apart from the interpolation rule defining the ODW weight schedule, all hyperparameters—including loss weights for curvature, normal consistency, and method-specific terms—are kept identical across experiments, matching the settings reported in the respective baselines. This controlled design isolates the effect of different interpolation strategies on reconstruction quality. We adopt keypoints $(s_0,w_0)=(0,10)$, $(s_1,w_1)=(0.2,10)$, $(s_2,w_2)=(0.5,0.001)$, and $(s_3,w_3)=(1.0,0)$ as tunable parameters empirically selected for all interpolation rules.
All experiments on an NVIDIA L4 GPU with 24 GB VRAM and the machine is equipped 32 GB of RAM.

\subsection{Evaluation Metrics}

We assess reconstruction accuracy using three standard metrics. Chamfer Distance (CD), scaled by $10^{3}$, quantifies the average discrepancy between two surfaces, where lower values indicate higher fidelity. F1 Score (F1) represents the harmonic mean of precision and recall, computed with a distance threshold of $5\times10^{-3}$ between predicted and ground-truth point sets. Results are scaled by $10^{2}$, with higher values reflecting better surface overlap. Normal Consistency (NC) measures the mean cosine similarity between predicted and reference normals, also scaled by $10^{2}$; higher values indicate stronger alignment. For all three metrics, we report the mean of all shapes in dataset subset.

\subsection{Qualitative Results}

\begin{figure*}[t]
    \centering
    \includegraphics[width=0.99\linewidth,trim= 150 0 150 0,clip]{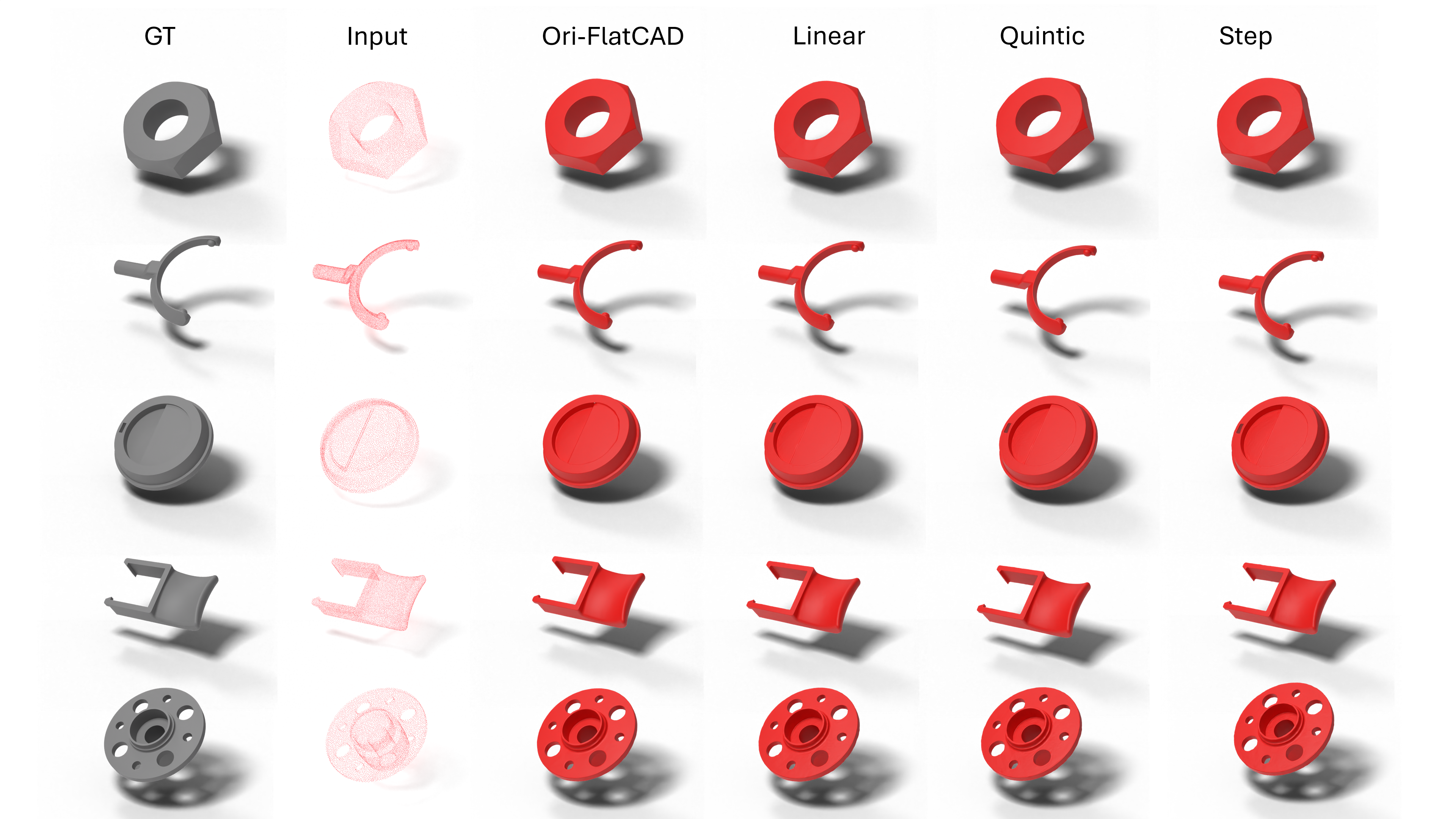}
    \caption{Comparison with the original FlatCAD~\cite{Yin2025FlatCAD}. We evaluate our proposed weight scheduling strategies—linear, quintic (fifth-order polynomial), and step interpolation—against the baseline with constant weights. All three schedules achieve comparable or superior reconstruction quality, consistently producing cleaner, more complete, and geometrically faithful surfaces that closely match the ground truth geometry (GT). The improvement arises from curvature-aware regularization that promotes developability, effectively suppressing spurious artifacts. For detailed numerical evaluation, please refer to the quantitative results reported in Table~\ref{tab:interpolation_compare}.}
    \label{fig:interpolation_compare}
\end{figure*}

Figure~\ref{fig:interpolation_compare} provides a visual comparison of reconstruction quality across different scheduling strategies. All methods are able to recover the overall shape reasonably well. Notably, in the third row (cup lid), the {FlatCAD} baseline fails to reproduce the drinking hole, resulting in an incorrect topology. In contrast, all scheduling variants (linear, quintic, and step) successfully capture this feature, demonstrating a clear advantage of time-varying sequencing for topologically sensitive reconstructions. This qualitative observation aligns with our quantitative results, highlighting that scheduling not only improves surface smoothness but also enhances the ability to recover correct topology.

\begin{table}[b]
  \centering
  \setlength{\tabcolsep}{5pt}
  \renewcommand{\arraystretch}{1.1}
  \small
  \caption{Quantitative results on the ABC dataset~\cite{Koch_2019_CVPR} comparing different sequencing methods. Evaluation is conducted on a subset using four metrics: Normal Consistency (NC), Chamfer Distance (CD), F1 score (F1) and pure training time. For each metric’s mean value, the best result is \underline{\textbf{bold underlined}}, and the second-best is \textbf{bold}. NC and F1 are reported $\times 10^2$, CD $\times 10^3$.}
  \label{tab:interpolation_compare}
  \begin{tabular*}{\linewidth}{@{\extracolsep{\fill}}lcccccccc}
    \toprule
     & \multicolumn{2}{c}{NC~$\uparrow$} & \multicolumn{2}{c}{CD$_{L_1}$~$\downarrow$} & \multicolumn{2}{c}{F1~$\uparrow$} & \multicolumn{1}{c}{time (s)} \\
    \cmidrule(lr){2-3} \cmidrule(lr){4-5} \cmidrule(lr){6-7}
     & mean & std & mean & std & mean & std & mean \\
    \midrule
    FlatCAD (w/o scheduling)   & 96.14 & 4.73 & 4.37 & 5.48 & 84.98 & 24.18 & \underline{\textbf{877.48}} \\
    Linear scheduling          & 97.95 & 1.60 & 3.05 & 2.17 & 90.59 & 16.89 & 882.65 \\
    Quintic scheduling         & \underline{\textbf{98.01}} & \underline{\textbf{1.46}} & \underline{\textbf{2.86}} & \underline{\textbf{1.22}} & \underline{\textbf{92.72}} & \textbf{10.08} & \textbf{878.21} \\
    Step scheduling            & \textbf{97.99} & \textbf{1.51} & \textbf{2.87} & \textbf{1.34} & \textbf{92.71} & \underline{\textbf{9.70}} & 1003.51 \\
    \bottomrule
  \end{tabular*}
\end{table}

\begin{table}[b]
  \centering
  \setlength{\tabcolsep}{5pt}
  \renewcommand{\arraystretch}{1.1}
  \small
  \caption{Quantitative results on the ABC dataset~\cite{Koch_2019_CVPR} comparing decreasing and increasing scheduling. Evaluation is conducted on a subset using four metrics: Normal Consistency (NC), Chamfer Distance (CD), F1 score and pure training time. For each metric’s mean value, the best result is \underline{\textbf{bold underlined}}, and the second-best is \textbf{bold}. NC and F1 are reported $\times 10^2$, CD $\times 10^3$.}
  \label{tab:increase_decrease_compare}
  \begin{tabular*}{\linewidth}{@{\extracolsep{\fill}}lcccccccc}
    \toprule
     & \multicolumn{2}{c}{NC~$\uparrow$} & \multicolumn{2}{c}{CD$_{L_1}$~$\downarrow$} & \multicolumn{2}{c}{F1~$\uparrow$} & \multicolumn{1}{c}{time (s)} \\
    \cmidrule(lr){2-3} \cmidrule(lr){4-5} \cmidrule(lr){6-7}
     & mean & std & mean & std & mean & std & mean \\
    \midrule
    FlatCAD (w/o scheduling)   & 96.14 & 4.73 & 4.37 & 5.48 & 84.98 & 24.18 & \textbf{877.48} \\
    Decreasing Linear scheduling & \underline{\textbf{97.95}} & \underline{\textbf{1.60}} & \underline{\textbf{3.05}} & \underline{\textbf{2.17}} & \underline{\textbf{90.59}} & \underline{\textbf{16.89}} & \ 882.65 \\
    Increasing Linear scheduling & \textbf{97.68} & \textbf{1.74} & \textbf{3.24} & \textbf{2.37} & \textbf{88.32} & \textbf{20.81} & \underline{\textbf{871.38}} \\
    \bottomrule
  \end{tabular*}
\end{table}

\subsection{Quantitative Results}
Table~\ref{tab:interpolation_compare} reports reconstruction performance on the ABC dataset across different scheduling strategies. Compared to the fixed-weight baseline FlatCAD~\cite{Yin2025FlatCAD}, all scheduling variants yield substantial improvements, confirming that time-varying weights consistently benefit optimization. Among them, quintic scheduling achieves the best overall performance and have CD imporves up to 35\% (cf. Table~\ref{tab:interpolation_compare}) . Overall, these results validate our hypothesis: interpolated decay schedules improve reconstruction quality, with quintic interpolation striking the best balance between stability, accuracy, and efficiency.

\subsection{Ablation on Scheduling Direction: Decay and Warm-Up}

\begin{figure*}[t]
    \centering
    \includegraphics[width=0.99\linewidth]{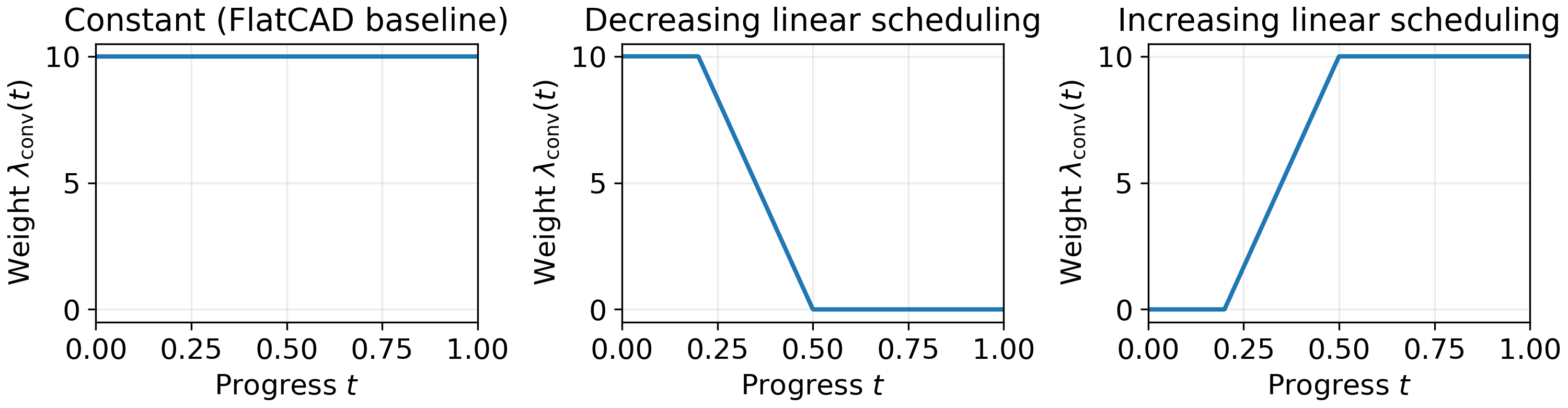}
    \caption{Scheduling strategies for the ODW weight $\lambda_{\mathrm{ODW}}(t)$. 
From left to right: Constant baseline (fixed $w=10$), Decreasing linear (strong--start/decay from $10$ to $0$), and Increasing linear (warm-up from $0$ to $10$). The horizontal axis shows normalized training progress $t \in [0,1]$.}
    \label{fig:decrease_increase}
\end{figure*}

To further validate the effectiveness of our proposed decay strategy, we conducted additional experiments designed to disentangle the contribution of the strong--start mechanism. Specifically, we compared our decreasing schedule against a symmetric increasing schedule (warm-up). The increasing schedule begins with $w=0$, holds this value for the first 20\% of training, then linearly ramps up to $w \approx 9.999$ by the halfway point, and finally reaches $w=10$ at convergence (see Fig.~\ref{fig:decrease_increase}). This setup mirrors the piecewise-linear structure of our decay scheme but reverses its direction. We benchmarked three cases: (1) the fixed-weight baseline of FlatCAD~\cite{Yin2025FlatCAD}, (2) our decreasing schedule (strong--start/decay), and (3) the increasing warm-up schedule. The quantitative comparison is summarized in Table~\ref{tab:increase_decrease_compare}.

\section{Discussion and Conclusions}

We presented scheduling strategies for the Off-Diagonal Weingarten (ODW) loss in neural SDF training for CAD models. 
Our experiments demonstrate that scheduling the Off-Diagonal Weingarten (ODW) loss substantially improves neural SDF training for CAD models. Among all variants, the decreasing schedule yields the most stable optimization and highest reconstruction fidelity—reducing Chamfer Distance by up to 35\% compared to FlatCAD. These results confirm that strong early curvature regularization prevents unstable minima, while gradual relaxation restores geometric detail. In contrast, increasing (warm-up) schedules fail to provide early stabilization and lead to noisier curvature. Overall, the benefit arises not merely from time variation but from the strong-start bias that guides learning toward globally consistent solutions. Future work will extend scheduling to other curvature- and topology-aware priors and investigate adaptive or data-driven weighting for broader shape categories.

\bibliographystyle{splncs04}
\bibliography{main}
\end{document}